\documentclass[aps,twocolumn,showpacs,preprintnumbers]{revtex4}
\newcommand{\nc}{\newcommand}
\nc{\rnc}{\renewcommand}
\nc{\beq}{\begin{equation}}
\nc{\eeq}{\end{equation}}
\nc{\bea}{\begin{eqnarray}}
\nc{\eea}{\end{eqnarray}}
\nc{\ba}{\begin{array}}
\nc{\ea}{\end{array}}
\nc{\nn}{\nonumber}
\nc{\bpi}{\begin{picture}}
\nc{\epi}{\end{picture}}
\nc{\scs}{\scriptstyle}
\nc{\ts}{\textstyle}
\nc{\unit}{{\mbox{\boldmath\large $1$}}}
\nc{\p}{\partial}
\nc{\ua}{\uparrow}
\nc{\da}{\downarrow}
\nc{\uada}{{\uparrow\downarrow}}
\nc{\f}{\frac}

\nc{\al}{\alpha}
\nc{\be}{\beta}
\nc{\ga}{\gamma}
\nc{\de}{\delta}
\nc{\la}{\lambda}
\nc{\si}{\sigma}
\rnc{\th}{\theta}
\nc{\Ga}{{\sf\Gamma}}
\nc{\La}{\Lambda}
\nc{\Si}{\Sigma}

\nc{\abar}{\bar{a}}
\nc{\J}{{\sf J}}
\nc{\T}{{\sf T}}
\rnc{\P}{{\sf P}}
\nc{\Q}{{\sf Q}}
\nc{\R}{{\sf R}}
\rnc{\S}{{\sf S}}
\nc{\Th}{\sf\Theta}
\nc{\U}{{\sf U}}
\nc{\V}{{\sf V}}
\nc{\Va}{{\sf V}_a}
\nc{\Vap}{{\sf V}_a'}
\nc{\Vb}{{\sf V}_b}
\nc{\Vc}{{\sf V}_c}
\nc{\Vd}{{\sf V}_d}
\nc{\X}{{\sf X}}
\nc{\Y}{{\sf Y}}
\nc{\Z}{{\sf Z}}

\nc{\cd}[2]{\scs\{\raisebox{-.46ex}{\rlap{\tiny{#2}}}
  \raisebox{.45ex}{\tiny{#1}}\scs\}}
\nc{\zm}{\ba{cc}0&0\\0&0\ea}
\nc{\zr}{\ba{cc}0&0\ea}
\nc{\zc}{\ba{c}0\\0\ea}

\begin{document}

\bibliographystyle{apsrev}

\title{Simplified Transfer Matrix Approach in the Two-Dimensional
Ising Model with Various Boundary Conditions}
\author{Boris Kastening}
\email[Email address: ]{ka@physik.fu-berlin.de}
\affiliation{Institut f\"ur Theoretische Physik\\
Freie Universit\"at Berlin\\ Arnimallee 14\\ D-14195 Berlin\\ Germany}

\date{24 May 2002}

\begin{abstract}
A recent simplified transfer matrix solution of the two-dimensional Ising
model on a square lattice with periodic boundary conditions is generalized
to periodic-antiperiodic, antiperiodic-periodic and antiperiodic-antiperiodic
boundary conditions.
It is suggested to employ linear combinations of the resulting partition
functions to investigate finite-size scaling.
An exact relation of such a combination to the partition function
corresponding to Brascamp-Kunz boundary conditions is found.
\end{abstract}

\pacs{05.50.+q, 68.35.Rh, 05.70.Ce, 64.60.Cn}

\maketitle

Recently, a solution of the two-dimensional Ising Model on a square
lattice with periodic boundary conditions was presented \cite{bo},
which simplifies Kaufman's version \cite{kaufman} of Onsager's solution
\cite{onsager} without requiring sophisticated mathematics.
Here we point out how the method easily generalizes also to the
cases of periodic-antiperiodic, antiperiodic-periodic and
antiperiodic-antiperiodic boundary conditions, which were recently
solved using a Grassmann path integral approach \cite{wuhululi}.
The motivation for varying the boundary conditions originates in the
investigation of finite-size scaling (FSS) \cite{wuhu,jake,izoghu}.
The analytic properties of FSS are of interest when trying to describe
critical properties of models whose exact solution is unknown.
For the analysis of FSS in solvable models it is very helpful if the
exact partition function can be written as a product rather than as a sum
\cite{jake,izoghu} as is the case with Brascamp-Kunz (BK) boundary conditions
\cite{brku}.
Here we exhibit linear combinations of partition functions with varying
boundary conditions that have this property and may be used for such an
analysis.
For one of these linear combinations we find an exact relation to the
partition function for BK boundary conditions.

We follow the notation in \cite{bo}.
Define the energy $E$ of the two-dimensional Ising model on a square
lattice with $m\times n$ sites by
\bea
-\be E
&=&
a\sum_{\nu=1}^n\left(\si_as_{m\nu}s_{1\nu}
+\sum_{\mu=1}^{m-1}s_{\mu\nu}s_{\mu+1,\nu}\right)
\nn\\
&&{}
+b\sum_{\mu=1}^m\left(\si_bs_{\mu n}s_{\mu1}
+\sum_{\nu=1}^{n-1}s_{\mu\nu}s_{\mu,\nu+1}\right),
\eea
where we assume $a,b>0$ as in \cite{bo} and where the spin variables
$s_{\mu\nu}$ can assume the values $\pm1$.
$(\si_a,\si_b)=(1,1)$ corresponds to periodic boundary conditions in both
directions (pp), as investigated in \cite{bo,kaufman,onsager}.
$(\si_a,\si_b)=(1,-1)$, $(-1,1)$ and $(-1,-1)$ correspond to
periodic-antiperiodic (pa), antiperiodic-periodic (ap) and
antiperiodic-antiperiodic (aa) boundary conditions, respectively.
The partition function is defined by
\beq
Z_{\al\be}^{(m,n)}(a,b)=\sum_{s_{\mu\nu}=\pm1}\exp(-\be E),
\eeq
where $\al\be$ may be pp, pa, ap or aa.
After defining $2^n\times2^n$ matrices $\X_\nu$ (and analogously $\Y_\nu$
and $\Z_\nu$) by
\beq
\X_\nu=
\underbrace{\unit\otimes\cdots\otimes\unit}_{\nu-1}\otimes\si_x\otimes
\underbrace{\unit\otimes\cdots\otimes\unit}_{n-\nu},
\eeq
where $\si_x$ is a Pauli matrix and $\unit$ is the $2\times2$ unit matrix,
we may write the partition function as
\beq
Z_{\al\be}^{(m,n)}(a,b)=[2\sinh(2a)]^{mn/2}\mbox{Tr}(\Q\V^m)
\eeq
where $\Q=\unit$ (now the $2^n\times2^n$ unit matrix) and
$\Q=\U_\X\equiv\X_1\X_2\cdots\X_n$ for periodic and antiperiodic boundary
conditions in the $a$ direction, respectively, and where
\beq
\V=\V_{a/2}\Vb\V_{a/2}
\eeq
with
\beq
\label{vahalf}
\V_{a/2}=\prod_{\nu=1}^n\exp(\abar\X_\nu/2),
\eeq
where $\abar$ is defined by $\sinh 2a\sinh2\abar=1$ as in \cite{bo}, and
\beq
\label{vb}
\Vb=\exp(\si_bb\Z_n\Z_1)\prod_{\nu=1}^{n-1}\exp\left(b\Z_\nu\Z_{\nu+1}\right)
\eeq
with $\si_b=1$ and $\si_b=-1$ for periodic and antiperiodic boundary
conditions in the $b$ direction, respectively.
With the same set of matrices
\beq
\J_{\al\be}^\pm=\P^\pm\J_{\al\be},~~~~~~
\P^\pm\equiv\frac{1}{2}(\unit\pm\U_\X)
\eeq
as in \cite{bo}, where
\bea
\J_{\al\be}
&=&
-\frac{i}{4}[\Ga_\al,\Ga_\be],
\\
\Ga_{2\nu-1}
&=&
\X_1\cdots \X_{\nu-1}\Z_\nu,
\\
\Ga_{2\nu}
&=&
\X_1\cdots \X_{\nu-1}\Y_\nu,
\eea
the partition function may be rewritten as
\beq
Z_{\al\be}^{(m,n)}(a,b)=[2\sinh(2a)]^{mn/2}\mbox{Tr}(\Q_\al\V_\be^m)
\eeq
with
\beq
\Q_{\cd{p}{a}}=\P_+\pm\P_-,
\eeq
\beq
\V_\be=\V_\be^+\V_\be^-=\V_\be^-\V_\be^+,
\eeq
\beq
\V_\be^\pm=\V_{a/2}^\pm\V_{b\be}^\pm\V_{a/2}^\pm,
\eeq
\beq
\label{vapm}
\V_{a/2}^\pm=\prod_{\nu=1}^n\exp(\abar\J_{2\nu,2\nu-1}^\pm),
\eeq
\beq
\label{vbp}
\V_{b\rm p}^\pm=\exp(\mp2b\J_{1,2n}^\pm)
\prod_{\nu=1}^{n-1}\exp(2b\J_{2\nu+1,2\nu}^\pm),
\eeq
\beq
\label{vba}
\V_{b\rm a}^\pm=\exp(\pm2b\J_{1,2n}^\pm)
\prod_{\nu=1}^{n-1}\exp(2b\J_{2\nu+1,2\nu}^\pm).
\eeq

From here on it is straightforward to follow the arguments given in
\cite{bo} to compute the partition function in terms of the $\ga_k$
defined by
\beq
\label{gak}
\cosh\ga_k=\cosh2\abar\cosh2b-\cos\frac{\pi k}{n}\sinh2\abar\sinh2b
\eeq
with the understanding that $\ga_k>0$ for $1\leq k\leq n$ and with
$\ga_0=2(\abar-b)$.
It turns out that antiperiodic boundary conditions in the $b$ direction
lead to an exchange of $\ga_k$ with even and odd indices, while
antiperiodic boundary conditions in the $a$ direction lead to a sign
change of half of the terms in the partition function, see (\ref{zap})
and (\ref{zaa}) below.

Define
\bea
\Si_{\rm ee}^{(m,n)}(a,b)
&=&
\sum_{\rm e}\exp\left(\frac{m}{2}\sum_{\nu=1}^n(\pm)\ga_{2\nu-2}\right),
\\
\Si_{\rm eo}^{(m,n)}(a,b)
&=&
\sum_{\rm e}\exp\left(\frac{m}{2}\sum_{\nu=1}^n(\pm)\ga_{2\nu-1}\right),
\\
\Si_{\rm oe}^{(m,n)}(a,b)
&=&
\sum_{\rm o}\exp\left(\frac{m}{2}\sum_{\nu=1}^n(\pm)\ga_{2\nu-2}\right),
\\
\Si_{\rm oo}^{(m,n)}(a,b)
&=&
\sum_{\rm o}\exp\left(\frac{m}{2}\sum_{\nu=1}^n(\pm)\ga_{2\nu-1}\right),
\eea
where the first index in $\Si_{\rm xy}$ and the index under the summation
sign refer to all combinations in the sum in the exponent that have an
even/odd number of minus signs and the second index of $\Si_{\rm xy}$
refers to even/odd indices of the $\ga_k$.
Define also
\bea
\label{codef}
C_{\rm o}^{(m,n)}(a,b)&\!=\!&[2\sinh(2a)]^{mn/2}
\prod_{k=1}^n\left[2\cosh\left(\frac{m}{2}\ga_{2k-1}\right)\right],
\nn\\\\
\label{sodef}
S_{\rm o}^{(m,n)}(a,b)&\!=\!&[2\sinh(2a)]^{mn/2}
\prod_{k=1}^n\left[2\sinh\left(\frac{m}{2}\ga_{2k-1}\right)\right],
\nn\\\\
\label{cedef}
C_{\rm e}^{(m,n)}(a,b)&\!=\!&[2\sinh(2a)]^{mn/2}
\prod_{k=1}^n\left[2\cosh\left(\frac{m}{2}\ga_{2k-2}\right)\right],
\nn\\\\
\label{sedef}
S_{\rm e}^{(m,n)}(a,b)&\!=\!&[2\sinh(2a)]^{mn/2}
\prod_{k=1}^n\left[2\sinh\left(\frac{m}{2}\ga_{2k-2}\right)\right],
\nn\\
\eea
and note that
\bea
\label{soid}
S_{\rm o}^{(2m,n)}(a,b)&=&S_{\rm o}^{(m,n)}(a,b)C_{\rm o}^{(m,n)}(a,b),
\\
\label{seid}
S_{\rm e}^{(2m,n)}(a,b)&=&S_{\rm e}^{(m,n)}(a,b)C_{\rm e}^{(m,n)}(a,b).
\eea
Then, omitting upper indices $(m,n)$ and arguments $(a,b)$ for brevity,
we have for the various cases
\begin{widetext}
\bea
\label{zpp}
Z_{\rm pp}
&=&
[2\sinh(2a)]^{mn/2}(\Si_{\rm eo}+\Si_{\rm oe})
=\f{1}{2}
(\phantom{+}C_{\rm o}+S_{\rm o}+C_{\rm e}-S_{\rm e}),
\\
\label{zpa}
Z_{\rm pa}
&=&
[2\sinh(2a)]^{mn/2}(\Si_{\rm ee}+\Si_{\rm oo})
=\f{1}{2}
(\phantom{+}C_{\rm o}-S_{\rm o}+C_{\rm e}+S_{\rm e}),
\\
\label{zap}
Z_{\rm ap}
&=&
[2\sinh(2a)]^{mn/2}(\Si_{\rm eo}-\Si_{\rm oe})
=\f{1}{2}
(\phantom{+}C_{\rm o}+S_{\rm o}-C_{\rm e}+S_{\rm e}),
\\
\label{zaa}
Z_{\rm aa}
&=&
[2\sinh(2a)]^{mn/2}(\Si_{\rm ee}-\Si_{\rm oo})
=\f{1}{2}
(-C_{\rm o}+S_{\rm o}+C_{\rm e}+S_{\rm e}).
\eea
\end{widetext}
The result for $Z_{\rm pp}$ is the one found in \cite{bo} and, observing
the different sign convention for $\ga_0$, in \cite{kaufman}.

In \cite{jake,izoghu}, FSS was investigated with BK boundary conditions
\cite{brku} so that the partition function has a product structure which
facilitates the subsequent analysis of the approach to the critical point
in the thermodynamic limit.
The partition functions above do not have such a product structure in
contrast to their linear combinations $C_{\rm o}$, $S_{\rm o}$, $C_{\rm e}$
and $S_{\rm e}$.
Except for $S_{\rm e}$, these combinations are positive for any temperature
and may be viewed as partition functions belonging to models with possibly
nonlocal boundary conditions.
Even though these boundary conditions may in general not be particularly
physical, the quantities $C_{\rm o}$, $S_{\rm o}$ and $C_{\rm e}$ may be
used as a mathematical tool to investigate the approach to criticality in
the same way as the partition function of the model with BK boundary
conditions.

$S_{\rm e}$ does not qualify for such an approach.
Since the pp boundary conditions cause a lower ground state energy than
the other three boundary conditions considered here, $S_{\rm e}$ becomes
negative for sufficiently low temperatures since $Z_{\rm pp}$ enters with
a minus sign.
$S_{\rm e}$ has another interesting property, though:
Due to the sign change of $\ga_0$ at the critical temperature of the
thermodynamic limit, (i.e., where $\abar=b$), $S_{\rm e}$ changes its sign
at the critical temperature for any size $m\times n$ of the square lattice.

The quantities (\ref{codef})-(\ref{sedef}) are the
same as in \cite{wuhu}, even though they have a different appearance.
In our notation, the expressions in \cite{wuhu} read
\begin{widetext}
\bea
\label{co}
C_{\rm o}^{(m,n)}(a,b)
&=&
2^{mn}\prod_{p=0}^{m-1}\prod_{q=0}^{n-1}
\left[\cosh(2a)\cosh(2b)
-\cos\f{(2p+1)\pi}{m}\sinh(2a)-\cos\f{(2q+1)\pi}{n}\sinh(2b)\right]^{1/2},
\\
\label{so}
S_{\rm o}^{(m,n)}(a,b)
&=&
2^{mn}\prod_{p=0}^{m-1}\prod_{q=0}^{n-1}
\left[\cosh(2a)\cosh(2b)
-\cos\f{2p\pi}{m}\sinh(2a)-\cos\f{(2q+1)\pi}{n}\sinh(2b)\right]^{1/2},
\\
\label{ce}
C_{\rm e}^{(m,n)}(a,b)
&=&
2^{mn}\prod_{p=0}^{m-1}\prod_{q=0}^{n-1}
\left[\cosh(2a)\cosh(2b)
-\cos\f{(2p+1)\pi}{m}\sinh(2a)-\cos\f{2q\pi}{n}\sinh(2b)\right]^{1/2},
\\
\label{se}
S_{\rm e}^{(m,n)}(a,b)
&=&
{\rm sgn}(1-\sinh(2a)\sinh(2b))\times
\nn\\
&&
2^{mn}\prod_{p=0}^{m-1}\prod_{q=0}^{n-1}
\left[\cosh(2a)\cosh(2b)
-\cos\f{2p\pi}{m}\sinh(2a)-\cos\f{2q\pi}{n}\sinh(2b)\right]^{1/2}.
\eea
\end{widetext}
While these expressions involve a double product instead of just a single
product as in (\ref{codef})-(\ref{sedef}), they have the advantage that
$C_o$ and $S_e$ are symmetric under the exchange $(a,m)\leftrightarrow(b,n)$
and that the only asymmetry of $C_e$ and $S_o$ under this exchange
originates in the boundary conditions and not in the mathematical treatment.

While Wu et al.\ \cite{wuhululi} use the Grassmannian integral approach
to map the problem onto a free fermion system, the method employed in this
work is close in spirit to Kaufman's solution \cite{kaufman}, but
avoiding some complications encountered there, as detailed in \cite{bo}.
While the Grassmannian method may be viewed as more elegant, our method is
accessible at a less sophisticated mathematical level.
It is also interesting to note the very different appearance of the solutions
(\ref{codef})-(\ref{sedef}) and (\ref{co})-(\ref{se}).
It would be interesting to see if our method can be generalized to cases
that so far have been solved only using the Grassmannian integral approach,
such as various boundary conditions on plane-triangular and honeycomb
lattices \cite{wuhu}, or other.

Brascamp and Kunz \cite{brku} gave a solution for a two-dimensional
Ising model on a square lattice with $(m+2)\times n$ sites with periodic
boundary conditions in the $b$ direction and fixed boundary conditions in
the $a$ direction.
If $k$ enumerates the sites in the $a$ direction ($0\leq k\leq m+1$), then
the spins at $k=0$ are fixed at $+1$, while the spins at $k=m+1$ are fixed
to alternate between $-1$ and $+1$, so $n$ must be even.
The resulting partition function was determined to be \cite{brku}
\begin{widetext}
\beq
Z_{\rm BK}^{(m,n)}(a,b)=2^{mn}\prod_{p=1}^m\prod_{q=1}^{n/2}
\left[\cosh(2a)\cosh(2b)-\cos\f{p\pi}{m+1}\sinh(2a)
-\cos\f{(2q-1)\pi}{n}\sinh(2b)\right].
\eeq
\end{widetext}
Using (\ref{soid}) and (\ref{co})-(\ref{ce}), we can establish the
connection
\bea
\label{sozbk}
S_{\rm o}^{(2(m+1),n)}(a,b)
&=&
S_{\rm o}^{(2,n)}(a,b)Z_{\rm BK}^{(m,n)}(a,b)^2
\nn\\
&=&
C_{\rm e}^{(n,2)}(b,a)Z_{\rm BK}^{(m,n)}(a,b)^2,
\eea
where the $S_{\rm o}$ may be factorized according to (\ref{soid}) and
$C_{\rm e}^{(n,2)}(b,a)$ most easily evaluated using (\ref{cedef}).

Eq.~(\ref{sozbk}) generalizes Eq.~(10) in \cite{izoghu} to the case $a\neq b$.
It would be interesting to understand (\ref{sozbk}) without having to resort
to the solutions of the models.
Then it may also be possible to relate $C_{\rm o}$, $S_{\rm o}$, $C_{\rm e}$
and $S_{\rm e}$ to yet other physical boundary conditions.

Investigation of FSS for $C_{\rm o}$, $S_{\rm o}$ and $C_{\rm e}$ is beyond
the scope of this paper.
Let us however state one interesting result that can be read off immediately
from (\ref{co}).
If $a=b$ and we define $z=\sinh2a$, the zeros $z_{pq}$ of $C_{\rm o}$
in the complex plane are given by $z_{pq}=\exp(i\al_{pq})$ with
\beq
\label{alpq}
\al_{pq}=\arccos\left[\f{1}{2}\left(\cos\f{(2p+1)\pi}{m}
+\cos\f{(2q+1)\pi}{n}\right)\right],
\eeq
and the zero nearest to the real axes is found for $p=q=0$.
For $m=n$, this zero scales as
\beq
\al_{00}=\f{\pi}{n}
\eeq
without any corrections as opposed to the case of BK boundary conditions
\cite{jake,izoghu}.
It remains to be investigated if simplifications may also be achieved
for the FSS of other quantities and if simplifying combinations
of partition functions can also be defined for models whose exact solution
is unavailable, e.g.\ by the use of symmetry arguments.

\end{document}